\newcommand\sovast{\ref@jnl{Soviet~Ast.}} 

\RequirePackage[normalem]{ulem} 
\RequirePackage{color}\definecolor{RED}{rgb}{1,0,0}\definecolor{BLUE}{rgb}{0,0,1} 

\documentclass[12pt,usenatbib]{mn2e}
\usepackage{url,ulem,times,graphicx,amsmath,amsfonts,amssymb,color,epsfig,epstopdf}
\usepackage{graphics}
\usepackage{hyperref}
\usepackage{epsf}
\usepackage{bm}
\usepackage{color}
\usepackage{rotating}
\usepackage[T1]{fontenc}
\usepackage{ae,aecompl}
 \usepackage{array,multirow}
\usepackage[percent]{overpic}

\def\lsim{\mathrel{\lower0.6ex\hbox{$\buildrel {\textstyle <}
 \over {\scriptstyle \sim}$}}}
\def\gsim{\mathrel{\lower0.6ex\hbox{$\buildrel {\textstyle >}
 \over {\scriptstyle \sim}$}}}

\def\eone{${\bf e}_{1}$}
\def\etwo{${\bf e}_{2}$}
\def\ethree{${\bf e}_{3}$}

\begin{document}

\title[Orientation of Planes in the QL Universe ]{The orientation of  planes of dwarf galaxies in the quasi-linear Universe}
\author[Libeskind et al]
{Noam I Libeskind$^{1,2}$, Edoardo Carlesi$^{1}$, Oliver M\"{u}ller$^{3}$, Marcel S Pawlowski$^{1,4}$, \newauthor Yehuda Hoffman$^{5}$, Daniel Pomar\`{e}de$^{6}$, Helene M Courtois$^{2}$, R. Brent Tully$^{7}$, \newauthor  Stefan Gottl\"{o}ber$^{1}$, Matthias Steinmetz$^{1}$,  Jenny Sorce$^{8,1}$, Alexander Knebe$^{9,10,11}$\\
$^1$Leibniz-Institut f\"ur Astrophysik Potsdam (AIP), An der Sternwarte 16, D-14482 Potsdam, Germany\\
$^2$University of Lyon; UCB Lyon-1/CNRS/IN2P3; IPN Lyon, France.\\
 $^3$Observatoire Astronomique de Strasbourg  (ObAS),
Universite de Strasbourg - CNRS, UMR 7550 Strasbourg, France\\
 $^4$Department of Physics and Astronomy, University of California, Irvine, CA 92697, USA; Hubble Fellow\\
$^5$Racah Institute of Physics, Hebrew University, Jerusalem 91904, Israel\\
$^6$Institut de Recherche sur les Lois Fondamentales de l'Univers, CEA Universit\'{e}. Paris-Saclay, 91191 Gif-sur-Yvette, France.\\
$^7$Institute for Astronomy (IFA), University of Hawaii, 2680 Woodlawn Drive, HI 96822, USA\\
$^{8}$Univ Lyon, Univ Lyon1, Ens de Lyon, CNRS, Centre de Recherche Astrophysique de Lyon
UMR5574, F-69230, Saint-Genis-Laval, France\\
$^{9}$Departamento de F\'isica Te\'{o}rica, M\'{o}dulo 15, Facultad de Ciencias, Universidad Aut\'{o}noma de Madrid, 28049 Madrid, Spain\\
$^{10}$Centro de Investigaci\'{o}n Avanzada en F\'isica Fundamental (CIAFF), Facultad de Ciencias, Universidad Aut\'{o}noma de Madrid, \\28049 Madrid, Spain\\
$^{11}$International Centre for Radio Astronomy Research, University of Western Australia, 35 Stirling Highway, Crawley, \\ Western Australia 6009, Australia\\
  }
\date{Accepted --- . Received ---; in original form ---}
\pagerange{\pageref{firstpage}--\pageref{lastpage}} \pubyear{2014}
\maketitle

 \begin{abstract}
To date at least 10 highly flattened planes of dwarf galaxies are claimed to have been discovered in the Local Universe.  The origin of these planes of galaxies remains unknown.  One suggestion is that they are related to the large-scale structure of the cosmic web. A recent study found that the normal of a number of these dwarf galaxy planes are very closely aligned with the eigenvector of the  shear tensor corresponding to the direction of greatest collapse obtained by reconstructing the full velocity field in the linear regime. Here we extend that  work by both considering an additional 5 planes beyond the 5 examined previously and by examining the alignment with respect to the quasi-linear field, a more sophisticated reconstruction, which is a better approximation on smaller (quasi-linear) scales. Our analysis recovers the previous result while not finding a significantly tight alignment with the additional 5 planes. However, the additional 5 planes normals also do not appear to be randomly oriented. We conclude that this could either be due to the normals of the new planes being poorly defined and described; the quasi-linear field at those locations being poorly constrained; or different formation mechanisms for the orientation of planes of dwarf galaxies.
\end{abstract}

\section{Introduction}
\label{section:intro} 
A current debate rages at conferences and in the literature regarding the existence of coherently moving planes of satellite galaxies in the Local Universe. On the one hand, scions in the field assert that explaining the existence of such planes may be the most important open problem in galaxy formation (Springel, private communication\footnote{This quote is attributed to V. Springel as he answered the question ``what is the most outstanding problem in galaxy formation?'' at 15th Potsdam Thinkshop {\it The role of feedback in galaxy formation} held in Potsdam from 3-7 September 2018. \url{https://twitter.com/satellitegalaxy/status/1036599442638610432}}). On the other hand some researchers question either the prevalence of such planes altogether \citep{2015MNRAS.453.3839P}, their non-transient nature \citep{2017MNRAS.465.641F,2018MNRAS.473.2212F}  or their alleged existential threat to the current dark matter, dark energy cosmological paradigm, known as the $\Lambda$CDM model \citep{2015MNRAS.452.3838C}.

\begin{table*}
 \centering
 \begin{tabular}{llrrrrrrcccc}
\hline
& Structure& $\hat{n}_{x}$ &$\hat{n}_{y}$ &$\hat{n}_{z}$& SG$_{x}$ &SG$_{y}$&SG$_{z}$&$\Delta$&$c/a$&$\sigma_{\hat{n}}$&Ref.  \\
\hline
\hline 
\\

&MW & 0.532&-0.306&-0.789& 0.000& 0.000& 0.000& 19.9 $\pm$ 0.3& $ 0.209 \pm 0.002$ &0.43$^o$&[1]\\  
&M31$_{\rm P1}$ &-0.339&-0.234& 0.912& 0.688&-0.303& 0.167&13.6 $\pm$ 0.2&0.107 $\pm$ 0.005& 0.79$^o$&[1]\\
&M31$_{\rm P2}$ &-0.108&-0.411& 0.905& 0.688&-0.303& 0.167&11.5&0.15 &N/P& [2] \\
&CenA$_{\rm P1}$ &-0.135&-0.442& 0.886&-3.410& 1.260&-0.330&73&0.2109$\pm$0.004& 2$^o$ & [3]\\
&CenA$_{\rm P2}$ & 0.079& 0.323&-0.943&-3.410& 1.260&-0.330&46&0.184$\pm$0.004&2$^o$ &[3]\\
&M101 & 0.629&-0.023&-0.778& 2.855& 5.746& 2.672& 46&0.03 & 1.5$^o$& [4]\\
&M83 &-0.654&-0.724& 0.221&-4.152& 2.601& 0.085&  20.4&0.097&N/P&[5]\\
&LG$_{\rm P1}$&  0.112&-0.278&-0.954& 0.186&-0.188&-0.109&54.8 $\pm$ 1.8& 0.077 $\pm$ 0.003& 0.41$^o$&[1]\\
&LG$_{\rm P2}$& -0.155&-0.729&-0.667& 0.148&-0.409& 0.610&65.5 $\pm$ 3.1&0.110 $\pm$ 0.004& 1.72$^o$&[1]\\
&GNP & 0.423&-0.438&-0.793&-0.050& 0.873&-0.700&  53.4 $\pm$ 1.5& 0.098 $\pm$ 0.004&0.6$^o$&[6]\\ 

\\
  \hline
  
 \end{tabular}
 \caption{Properties of the 10 planar structures examined in this paper. From left to right we present the name of the planar structure (or the name of the galaxy around which it is found); the $x,y,z$ directions, in supergalactic coordinates of the unit normal, $\hat{n}_{i}$, to the plane; the $x,y,z$ positions in supergalactic coordinates of the centroids, SG$_{i}$, of each plane (in units of Mpc); the rms thickness $\Delta$, of the planar structure (in kpc); the ratio of the short to long axis, $c/a$ of the identified planar structure; and the error on the published normal direction ($\sigma_{\hat{n}}$), N/P means that no error was published on this normal. References for these are: [1]  \citealt{2013MNRAS.435.1928P}, [2] \citealt{2013MNRAS.436.2096S}, [3] \citealt{2015ApJ...802L..25T}, [4] \citealt{2017A&A...602A.119M}, [5] \citealt{2018A&A...615A..96M}, [6] \citealt{2014MNRAS.440..908P}
 }
\label{tab:planes}
\end{table*}

Debate not withstanding, dwarf galaxy planes -- some coherently moving, others identified only via the positions of dwarf galaxies -- have been shown or claimed to exist throughout the Local Universe. Specifically: one plane around the Milky Way \citep{1976RGOB..182..241K,1976MNRAS.174..695L,LyndenBell1982,2005A&A...431..517K,MetzKroupaLibeskind2008,2009MNRAS.394.2223M,2013MNRAS.435.2116P,2013MNRAS.435.1928P}, two around M31 \citep{2013Natur.493...62I,2013ApJ...766..120C,2013MNRAS.436.2096S}, two around Centaurus A (\citealt{2015ApJ...802L..25T,2018Sci...359..534M} but see also \citealt{2016A&A...595A.119M}), one around M101 \citep{2017A&A...602A.119M}, and one around M83 \citep{2018A&A...615A..96M}. \cite{2013MNRAS.435.1928P} took an alternative approach and found two vast planar structures by searching for preferential planar alignments among the positions of field dwarfs in the Local Group, naming these ``Local Group plane 1 and 2'' (LG$_{\rm P1}$, LG$_{\rm P2}$ for short). Using a similar approach \cite{2014MNRAS.440..908P} identified a large planar structure called ``The Great Northern Plane'' (GNP, since it is in the northern galactic hemisphere), by associating NGC3109 with the nearby dwarfs around the Milky Way. A summary of the properties of these planes is shown in Table.\ref{tab:planes}. Note that some dwarf galaxies in the Local Group are associated with more than one of these structures, such that not all planes are parallel or strictly independent (some may be intersecting).

\citet{2014Natur.511..563I} attempted the first statistical investigation of the prevalence of co-rotating planes of satellite galaxies using pairs of diametrically opposed satellites in SDSS. They report a potentially high incidence of such structures (up to 60 per cent), but their results have since been questioned (\citealt{2015MNRAS.453.3839P}, \citealt{2015MNRAS.449.2576C}, but see \citealt{2015ApJ...805...67I} for a response.)

The origin of these planes is a matter of debate and remains an open question. \cite{2018MPLA...3330004P} presents a comprehensive review not only of all the properties of these planes, but of the pros and cons of the  three possible scenarios put forward to explain their origins: filamentary accretion, group accretion, or tidal dwarfs. The filamentary accretion hypothesis recently found some support as \cite{2015MNRAS.452.1052L} observed that four of these satellite planes (the two planes around M31 and Centaurus A) are directly aligned with, and thus likely related to, the local large-scale structure. \cite{2015MNRAS.452.1052L} used a Wiener filter reconstruction of the cosmic density field from the ComicFlows-2 survey \citep{2013AJ....146...86T} to compute the principal directions along which matter is compressed. Specifically, the claim was made that principal axes of the cosmic web as characterised by the tidal tensor, are aligned with these planar structures, an alignment consistent with theoretical studies on the geometry of satellite accretion \citep{2014MNRAS.443.1274L,2015ApJ...813....6K}. This hypothesis would suggest that the large scale tidal field endows these satellite planes with their orientation and thus must play a causal role in their formation thereby coupling small non-linear scales with larger linear and quasi-linear (QL) scales.

The work of \cite{2015MNRAS.452.1052L} is extended here in two different aspects. First, new dwarf galaxy planes have been discovered since \cite{2015MNRAS.452.1052L} namely around M101 and M83. We also add to our analysis the three large structures of dwarf galaxies in Local Group (the LG$_{\rm P1}$, LG$_{\rm P2}$ and the GNP). Where \cite{2015MNRAS.452.1052L}  examined the alignments of the planes with respect to the cosmic web defined by the linear Wiener filter reconstruction of the velocity shear tensor here the local cosmic web is evaluated by means of the QL reconstruction of the density field in the Local Universe \citep{2018NatAs...2..680H}. Given that the QL reconstruction is a better approximations of the shear field on scales smaller than the linear Wiener filter, we expect to gain insight in the relationship between of dwarf galaxy planes and QL scales.

\section{Methodology}
\subsection{Observed dwarf planes}
There is little doubt that planar structures composed of small dwarf galaxies, have been observed throughout the Local Universe. However, it would be folly to assert that the cornucopia of locally observed dwarf galaxy planes represent a single astronomical class of object. This is because observations of each dwarf galaxy plane differ in important ways. For example on the one hand, around the Milky Way we have proper motion measurements \citep{MetzKroupaLibeskind2008,2018arXiv180500908F}  which indicate that the satellite galaxies co-rotate within the plane they define.  For M31 and Centaurus A, no proper motions exist, yet sub samples of the dwarfs around those galaxies do show line-of-sight motion consistent with co-rotation (or a shear).  A fair fraction of the Milky Way's halo is obscured due to the galactic disk and the zone of avoidance -- not a significant issue when imaging the surroundings of other galaxies. Despite the fact that the non-uniform sky  coverage has led some to assert a bias in detecting dwarf galaxy planes, \cite{2016MNRAS.456..448P} has demonstrated this is a minor issue. Furthermore, due to magnitude limits at the faint end of the luminosity function, only the brighter dwarfs are observed around more distant galaxies like M101 or M83 than in the Local Group.  Such dwarf galaxy surveys probe the dwarf galaxy population in the regime of the ``classical'' dwarfs ($M_V<-10$ mag) due to the expensive, time intensive observations needed to identify lower surface brightness objects \citep[e.g][]{2017A&A...602A.119M,2016ApJ...823...19C}. Typically, distances are estimated by resolving stellar populations and measuring the magnitude of stars at the tip of the red giant branch \citep[e.g.][]{2001A&A...371..487J,2013AJ....145..101K,2018A&A...615A..96M,2019arXiv190702012M}. However such measurements are subject to 5-10\% errors, which at large distances, renders plane estimates erroneous (unless, of course, the plane is viewed edge on such that errors scatter galaxies within an observed plane, as in M101 or Centaurus A; see Fig. 7 in \citealt{2017A&A...602A.119M}). Lastly planar structures may be composed of satellite galaxies or field dwarfs namely found on sub-virial radius scales or on Mpc scales. For example the planes found around the MW or  M31$_{\rm P1}$ are composed  of satellite galaxies, while others (such as GNP, LG$_{\rm P1}$ and LG$_{\rm P2}$) are explicitly defined as being composed of non-satellite dwarfs (although these do not exclude backsplash galaxies which were at one point within the virial radius of their host). Finally some of the planes (such as the plane of M101 dwarfs, CenA$_{\rm P1}$ and CenA$_{\rm P2}$) are composed of both dwarfs in and outside of their host's virial radius \citep[where the virial radius is roughly defined based on an assumed model, e.g. see Fig 7 in][]{2018AJ....156..105A}.

Therefore it is with these words of caution that we examine here the 10 known structures published in the literature and presented in Table~\ref{tab:planes}. For the purpose of this paper we maintain a neutral approach towards any controversy surrounding the quantification of each plane, referring the reader to the literature for specific definitions of the plane normal.

\subsection{The quasi-linear reconstruction of the local density field}
\label{sec:WF}
In this section we describe how the underlying density field of the local Universe is obtained. We note that this section is, broadly speaking, a review of the work of \cite{2018NatAs...2..680H} and we refer the reader to that paper for details on the reconstructions method, limitations and application.

According to the theory of gravitational instability \citep{1970A&A.....5...84Z} the growth of structure is accompanied by the acceleration of matter from under dense regions towards over dense regions. Before structure formation becomes non-linear the density field, and the velocity field it engenders, are related by a simple convolution. Therefore, when confined to linear scales, observations of the peculiar velocity field can be used to infer the underlying density field, the source of the observed velocities. Such an approach, termed `constrained realisations' has a long history in the literature starting with the seminal work of \cite{1987ApJ...323L.103B}  and extended by the Hoffman-Ribak  algorithm \citep{1991ApJ...380L...5H}. Noting that the peculiar velocity field is noisy and sparsely sampled, \cite{1999ApJ...520..413Z} introduced the Wiener filter method for producing reconstructions: a deconvolution of the velocity field in the presence of noise. Given that according to the $\Lambda$CDM cosmological model the universe is composed of Gaussian random fields, the Wiener filter is ideal since it results in a minimal variance estimate of the reconstructed field. 

The Wiener filter results in a reconstructed field which depends on the underlying unconstrained (random) scales. Therefore many constrained realisations can be constructed from one observed peculiar velocity field to which the Wiener filter has been applied. These constrained realisations can be used to construct constrained initial conditions for N-body simulations  \citep[e.g. see][]{2016MNRAS.458..900C,2016MNRAS.455.2078S}

In practice we start with the publicly available peculiar velocity catalog known as CosmicFlows-2 \citep{2013AJ....146...86T}. These are then grouped so that the virial motions of galaxies that are gravitationally bound together, do not count as individual tracers  of the peculiar velocity field. Instead a weighted mean based on the distance modulus of each group member is computed, such that a single ``mean''  peculiar velocity may be assumed for the group or cluster \citep{2013AJ....146...86T}. After the Wiener filter and constrained realisation algorithm  described above is applied, constrained initial conditions for $N$-body simulations are produced.

In order to construct the QL maps of \cite{2018NatAs...2..680H}, some 20 such constrained $N$-body simulations are run. Note that as shown in \cite{2018NatAs...2..680H} 20 constrained simulations is sufficient to obtain stable, converged results with respect to QL scales. These are periodic boxes of co-moving side length $L_{\rm box} =500$ Mpc$/h$, co-moving softening of 25~kpc$/h$ and which follow the evolution of 512$^{3}$ particles using the publicly available GADGET-2 code \citep{2005MNRAS.364.1105S} assuming Planck-1 cosmological parameters (i.e. $\Omega_{\rm m}=0.31$, $\Omega_{\Lambda}=0.69$, $H_{0}=67.7$km/s/Mpc, $\sigma_{8}=0.83$).

A QL estimate of the velocity and density field, is obtained by taking the arithmetic and geometric mean (respectively) of these 20 constrained simulations. The averaging process effectively smooths out any non-linearities resulting in a QL estimate. Note that this process does well in uncovering structures in the QL density field, and hence we focus on the tidal tensor (described below), the Hessian of the gravitational potential of the reconstructed density field. We note that in the linear regime, the tidal tensor converges with the velocity shear tensor (namely, in the linear regime the velocity shear is directly engendered by the tidal field).  \cite{2018NatAs...2..680H} showed that the QL velocity field is very close to the linearly reconstructed velocity field on large scales - they differ on small scales due to non-linearities. Therefore we opt to focus on the QL tidal tensor rather than the velocity shear tensor.

\subsection{Tidal Tensor}

The tidal tensor, inspired by \cite{1970A&A.....5...84Z} and applied to $N$-body simulations by \cite{2007MNRAS.375..489H}, is based on the Hessian of the gravitational potential ($\phi$), and is defined as
\begin{equation}
T_{\alpha\beta}=\frac{\partial^{2}\phi}{\partial x_{\alpha}\partial x_{\beta}}
\end{equation}
where the gravitational potential is normalised by $4\pi G\bar{\rho}$ (where $G$ is Newton's gravitational constant, $\bar{\rho}$ is the mean density of the universe and $\delta$ is the matter over density)  such that the potential satisfies the Poisson equation $\nabla^{2}\phi=\delta$ in co-moving coordinates (note that since we are dealing with local planes the expansion factor, $a=1$, is effectively omitted).

\begin{figure*}
 \includegraphics[width=40pc]{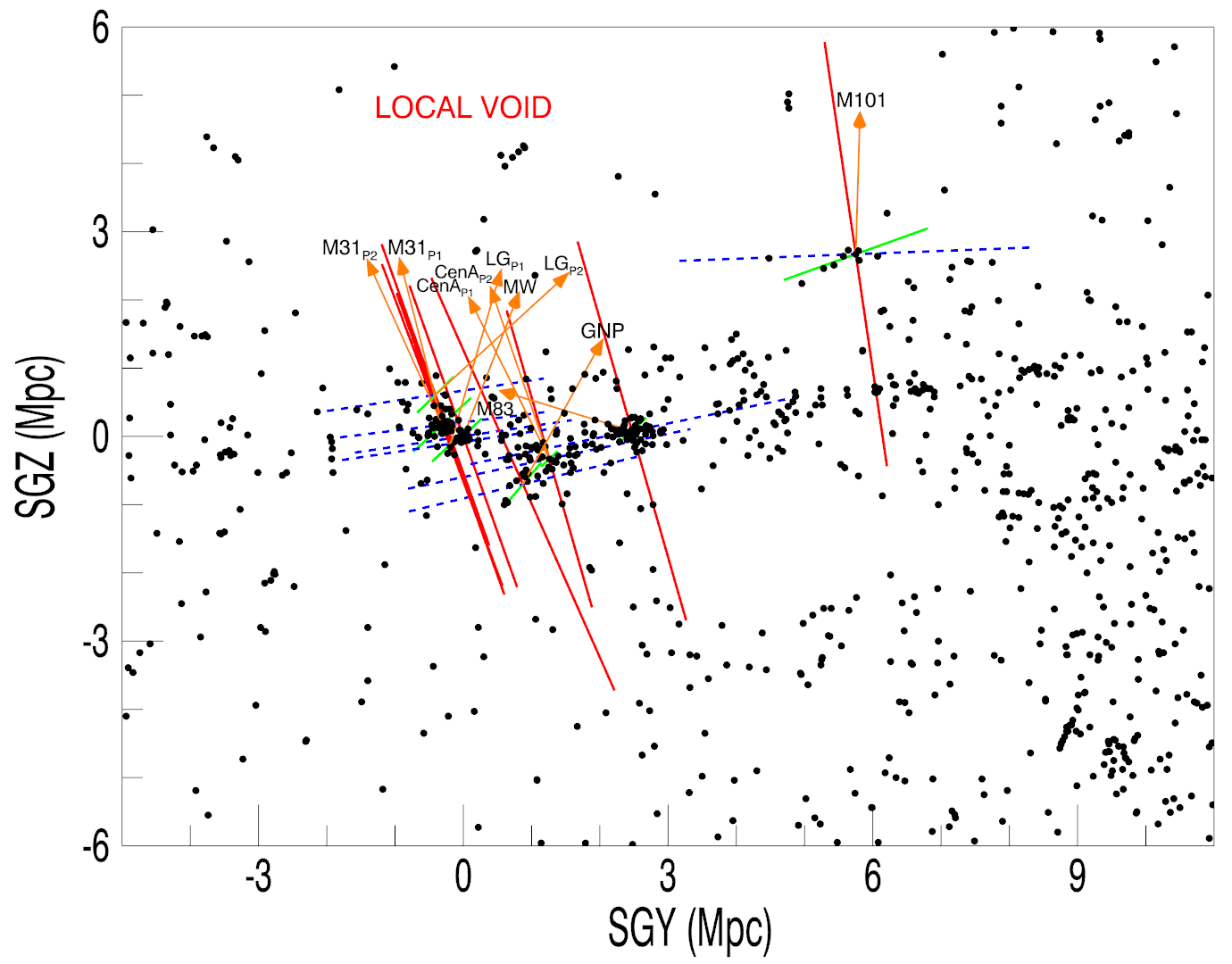}
\caption{A map of the Local Universe in supergalactic coordinates, considered in this work. The normal of each plane is denoted by the orange arrows and its eigenframe (\eone, \etwo~and \ethree), measured at each plane's putative center, is shown by the red, green and blue lines. Solid lines denote positive eigenvalues  (associated with compression) while dashed lines reference negative eigenvalues, (associated with dilatation). Black dots are Local Universe galaxies from the \citealt{2017ApJ...843...16K} catalog. Plane names are added as black labels along with the Local Void. This figure is accompanied by an \href{https://sketchfab.com/3d-models/cf2-ql-eigenvectors-planes-localsheet2virgo-v002-05a270e6b97a467cb67878cff7a59dde}{interactive graphic.}} \label{fig:cosmography}
 \end{figure*}

\begin{figure}
 \includegraphics[width=20pc]{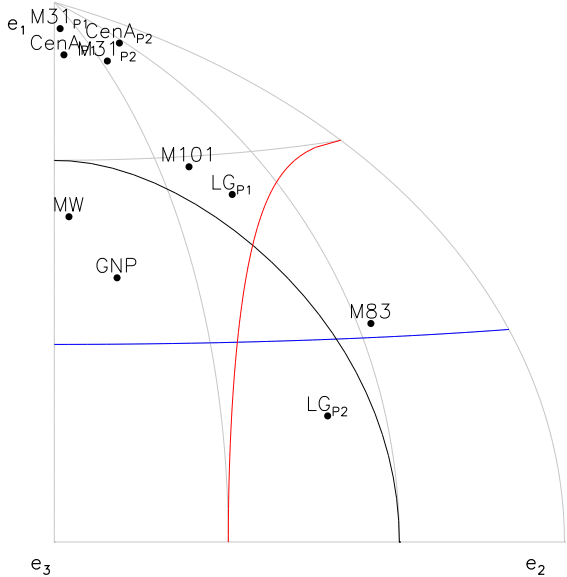}
\caption{The location of the normals of the 10 structures considered here with respect to the orthonormal eigenvectors of  the tidal tensor computed at the location of each plane's centroid. Since the three orthonormal eigenvectors have an arbitrary direction, they constitute an octant of cartesian space. The normals  of the structures considered here are shown on this octant in order to clarify the orientation of the planes with respect to the shear eigenvectors. In blue, red, and black we show regions 60 degrees away from \eone, \etwo, and \ethree, respectively. A random distribution of points on such an octant would have  the same number of points within and beyond 60 degrees. }
 \label{fig:octant}
 \end{figure}

\begin{figure}
 \includegraphics[width=20pc]{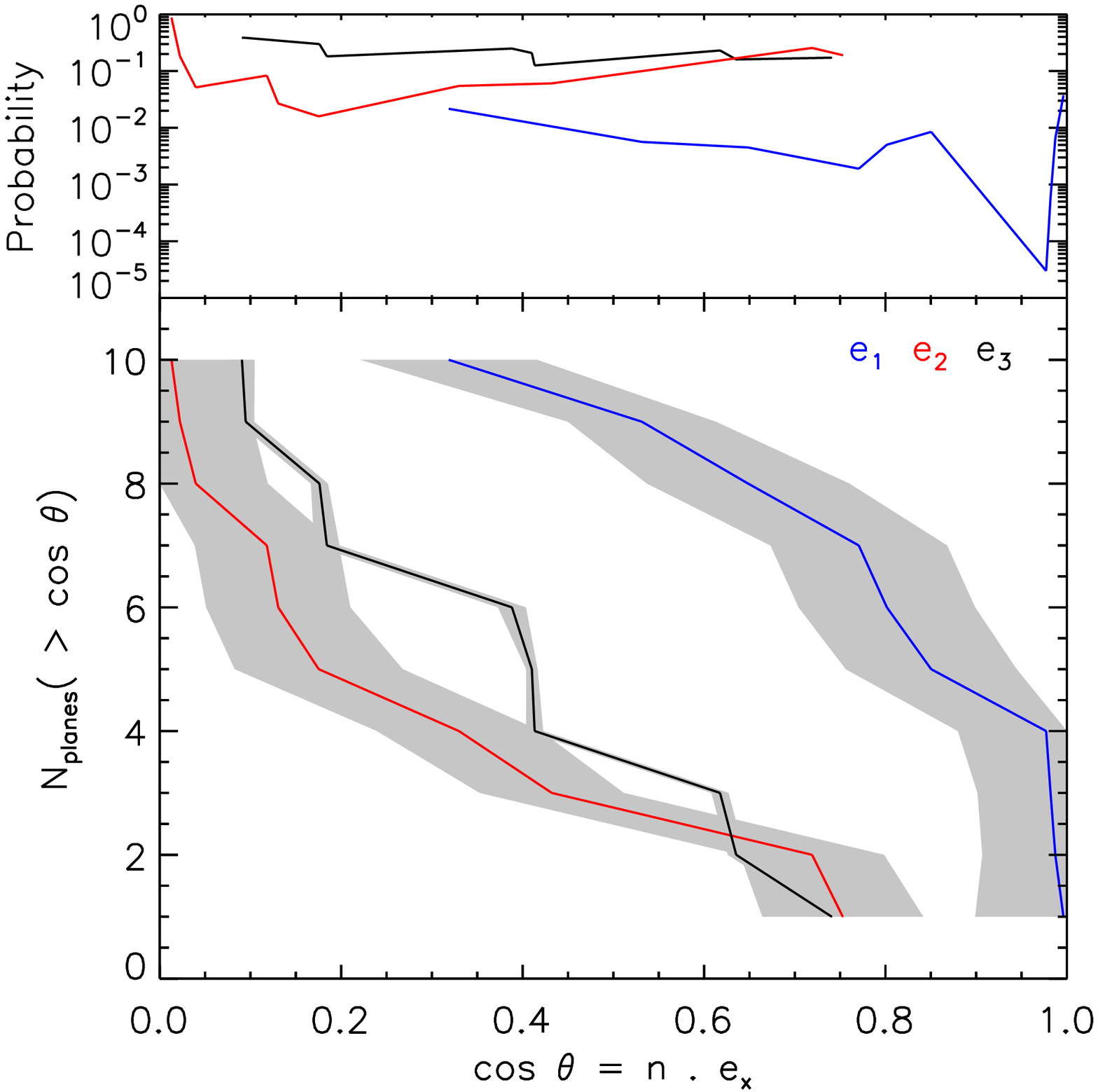}
\caption{Bottom: The cumulative distribution function of the angle made between each plane normal and the three eigenvectors of the tidal tensor (\eone, \etwo, and \ethree~in blue, red and black respectively). The shaded grey region represents the standard deviation in the direction of the eigenvectors of the 20 individual constrained simulations which are used to construct the QL reconstruction. Top: the probability of randomly drawing N$_{\rm planes} (> \cos\theta)$ (out of a sample of 10) from a uniform distribution.}
 \label{fig:stat}
 \end{figure}

In order to compute the principal direction of the cosmic web, namely the eigenvectors of the tidal  tensor, in the QL regime, we begin by constructing a $512^{3}$ Cloud-in-Cell mesh of each of the 20 $z=0$ simulation snapshots. Given the box size this results in a grid cell size of $\approx1$Mpc/h. This sets the formal scale of the tidal tensor's resolution. A further Gaussian smoothing with kernel size equivalent to two cells is done (by a Fast Fourier Transformation in $k-$space) in order to eliminate any preferential directions spuriously introduced by the imposition of the (Cartesian) clouds-in-cell (CIC)  grid. This results in an  effective resolution of $\sim 2$Mpc/h. These 20 density fields are then (geometrically) averaged. The geometric mean, as opposed to the arithmetic mean, is taken because the density field has a log-normal distribution. The tidal tensor (equation 1) is then computed from the averaged density field and then diagonalised at the location of each cell.  Because of the degeneracy of the orthonormal eigenvectors with respect to direction mentioned above, the `eigenframe' defined by the diagonalised tidal tensor constitutes a single octant of three dimensional Cartesian space.

The normals to the ten planar structures identified in the Local Universe are then compared to the eigenvectors of the tidal tensor computed at the location of their centroids (see discussion at the end of section 3.1). 

\section{Results}
\subsection{Plane alignments with the eigenframe}
We begin the presentation of our results with a cosmographic map, Fig.~\ref{fig:cosmography} of the Local Universe in super-galactic coordinates. The normal to each plane is shown as well as the directions of the three eigenvectors. Fig.~\ref{fig:cosmography} shows the general tendency for these satellite plane normals to point towards the local void \citep{1987ang..book.....T,2019arXiv190508329T} or \eone, the axis along which material is being compressed fastest. We now quantify these orientations.

In Fig.~\ref{fig:octant} where we show the location of the 10 normals with respect to the QL reconstructed tidal eigenframe at their respective centroid positions (black symbols). That the two planes around M31 (M31$_{\rm P1}$ and M31$_{\rm P2}$) and the two planes around Centaurus A (CenA$_{\rm P1}$ and CenA$_{\rm P2}$) are located close to \eone~ was first noted by \cite{2015MNRAS.452.1052L}, as was the $\sim40$deg offset of the Milky Way's plane of satellites. These alignments are confirmed here by the QL tidal field. The additional 5 planes considered in this work (M101, M83, LG$_{\rm P1}$, LG$_{\rm P2}$ and GNP) do not lie as close to \eone~as the others, although the plane around M101 lies slightly closer than the plane around the Milky Way. The angle between the plane normal and the QL eigenvectors (as well as the error computed in  Section 3.2) is shown in Table.~\ref{tab:align_err}

With the exception of  LG$_{\rm P2}$, the additional planes considered have $\cos\theta >0.5$ or are within of 60 degrees of \eone. Combined with the other 5 planes around the Milky Way, M31 and Centaurus A, we therefore find that just 1 out of 10 plane normals to be inclined by more than 60 degrees, the outlier being LG$_{\rm P2}$. Similar, but slightly weaker statements can can be made regarding the avoidance of \etwo~and \ethree~ (with 7/10 and 8/10 plane normals in the $\cos\theta < 0.5$ region, respectively). The reader will note that the hypothesis that the normals are randomly oriented with respect to the tidal eigenframe would predict a uniform distribution in $\cos\theta$ with roughly half the planes having angles greater or less than 60 degrees. We investigate the probability of such a set up being random  below. 

In the bottom panel of Fig.~\ref{fig:stat} we show the (cumulative) number of planes ($\rm N_{\rm planes}$) whose normal is inclined by more than some $\cos\theta$ with respect to the eigenvectors \eone, \etwo, or \ethree~(in blue, red, and black, respectively). The propensity for plane normals to align with \eone~is seen quite clearly. For example 7 out of 10 planes are within $\cos\theta \approx 0.75$ ($\approx$ 40 degrees) of \eone, while none are within this angle of \etwo~or \ethree. In grey we show the error obtained (see below, section 3.1).

The probability of finding such angular distributions of normals is computed by randomly drawing 10 numbers, many (say 1,000,000) times, uniformly distributed between  0 and 1 and asking what fraction follow the cumulative distributions, namely have at least $\rm N_{\rm planes} (< \cos\theta)$. This is shown in the top panel of Fig.~\ref{fig:stat}. The fact that the most aligned four planes have a probability of occurring randomly of roughly 1:100,000 can be clearly seen. Since the next best aligned plane, that of M101, is tilted away from \eone~by roughly 30 degrees, this is enough to significantly increase the probability of a random draw to around 1:100. The orientation of the remaining planes with respect to \eone~are expected to occur randomly around the 1:100 to 1:1000 range while the full set of 10 has a probability of occurring randomly of a few percent. This is to be contrasted with the orientations with respect to \etwo~and \ethree~which are more or less consistent with random draws, only occasionally falling to the percent level. However, we caution that the planes tested and described in this work are not strictly (statistically) independent entities. Namely the existence of M31$_{\rm P2}$ is constructed entirely out of the dwarfs around M31 but excluded from M31$_{\rm P2}$. Similarly  the existence of LG$_{\rm P1}$ and LG$_{\rm P2}$ are not independent of the dwarf galaxies they contain. Therefore the statistical measure provided here should be taken as a rough estimate for this probability.

In the paragraph above (and in the top panel of Fig.~\ref{fig:stat}), each eigenvector is treated independently and we ask {\it how likely is a given distribution of normals with respect to a given eigenvector, \eone, \etwo~or \ethree?} We could equally ask the stricter conditional question: {\it how likely is a given distribution of normals with respect to all three eigenvectors, \eone, \etwo~and \ethree~simultaneously?} In this case the chance of finding a set of random numbers distributed in a way consistent with the plane normals is around 0.0006\%. 

A  note on the non alignment of the plane normals with the eigenframe is in order at this time. The centroids of the planes of dwarf galaxies considered here are located within a relatively small volume of the Local Universe,  with the most distant plane (M101) being $\sim7$Mpc away (most of the centroids of the planes examined here are located $\sim1$Mpc away). Throughout this volume the directions of the QL eigenframe is stable; the eigenvectors at the position of each centroid are roughly parallel. Specifically the mean angle between each pair of  \eone, \etwo~ and \ethree~vectors is 6.5$^{o}$, 8.8$^{o}$, and 6.7$^{o}$, respectively. The normals of the planes on the other hand are by no means correlated or parallel: the mean angle subtended  by each pair of plane normals is 41.9$^{o}$. This indicates that the ``blame'' for the non-existence of a tight correlation between all plane normals and the QL eigenframe lies with the plane normals, not the QL eigenframe.

\begin{table}
 \centering
 \begin{tabular}{llll}
\hline
 Structure& $\hat{n}~\cdot~$\eone& $\hat{n}~\cdot~$\etwo& $\hat{n}~\cdot~$\ethree\\
\hline
\hline 
\\
MW&$0.771 \pm 0.097$ & $0.040 \pm 0.079$ & $0.635 \pm 0.009$\\
M31$_{\rm P1}$&$0.996 \pm 0.097$ & $0.013 \pm 0.079$ & $0.095 \pm 0.009$\\
M31$_{\rm P2}$&$0.977 \pm 0.097$ & $0.118 \pm 0.079$ & $0.176 \pm 0.009$\\
CenA$_{\rm P1}$&$0.982 \pm 0.080$ & $0.022 \pm 0.079$ & $0.184 \pm 0.014$\\
CenA$_{\rm P2}$&$0.987 \pm 0.080$ & $0.131 \pm 0.079$ & $0.091 \pm 0.014$\\
M101&$0.850 \pm 0.094$ & $0.330 \pm 0.090$ & $0.410 \pm 0.006$\\
M83&$0.531 \pm 0.082$ & $0.753 \pm 0.089$ & $0.388 \pm 0.015$\\
LG$_{\rm P1}$&$0.802 \pm 0.097$ & $0.432 \pm 0.079$ & $0.413 \pm 0.009$\\
LG$_{\rm P2}$&$0.319 \pm 0.097$ & $0.719 \pm 0.079$ & $0.618 \pm 0.009$\\
GNP&$0.649 \pm 0.112$ & $0.175 \pm 0.093$ & $0.741 \pm 0.009$\\

  \hline
 \end{tabular}
 \caption{ The alignment (expressed as the cosine of the angle subtended) between each plane normal and each of the three orthonormal  eigenvectors of the tidal tensor of the reconstructed quasi-linear field at the location of the plane centroid. The errors are computed as the standard deviation of the eigenvectors at the same location for the 20 constrained simulations that are used in the quasi-linear reconstruction.}
\label{tab:align_err}
\end{table}

\subsection{Error Estimate} In the $\Lambda$CDM paradigm the over-density field (and thus the generated velocity field) is described by a Gaussian random field \citep{1986ApJ...304...15B}. The value of the density and velocity field at each given point is  thus a random Gaussian variable with a mean of zero. The QL reconstruction of the density field is a geometric mean of the the constrained simulations: the geometric (and not arithmetic) mean is taken because the density field has a log-normal distribution. The QL tidal field and its eigenvectors, computed from the average field, are thus not expected to be equivalent to the average of the eigenvectors computed for each constrained simulation. However, the variance of these gives us a feeling for how well constrained the direction of the QL eigenvectors are.

An error can thus be computed by examining the standard deviation of the tidal tensor eigenvectors' direction at each of the 10 locations considered. The alignment angle between the plane normal and the QL eigenvectors, as well as the error computed in this way is shown in Table.~\ref{tab:align_err}. The errors are on the  order of 0.1 in the cosine for \eone. The error is around an order of magnitude less for the direction of \ethree~making this the most stable direction; as has been noted before the filamentary direction towards Virgo along which \ethree~points, is one of the most stable aspects of local reconstructions based on velocities \citep[e.g][]{2016MNRAS.458..900C}. The errors are large enough that on the one hand, if we are generous, 8 out of 10 planes could be inclined by less than $\approx 40$ degrees with respect to \eone. On the other hand if we are pessimistic in the extreme, it means that no plane normal is closer than $\approx 25$ degrees to \eone.

We note that our error estimate establishes \ethree~as the most well constrained eigenvector. In fact this is consistent with our previous work which shows that the filamentary direction toward virgo as being the most robust feature of the local cosmography \citep{2016MNRAS.458..900C}.

\section{Conclusions and Discussion}
The origin of planes of dwarf galaxies  remains an open, controversial problem in cosmology. Such planes, observed in the Local Universe, are effectively defined by having a short to long  axis ratio of something small, typically of around 0.1. To date, no single theory dominates the debate, properly explaining all of the data.  One of the reasons for this is  that the planes that have been observed do not share identical characteristics. The numbers and luminosity of the dwarfs defining them varies as do the observational biases and the availability of either proper motions or accurate distances. It is thus unclear from the outset that the planes represent a single class of astronomical object - this remains to be seen and is beyond the scope of this work. On the other hand their existence remains firmly established.

As such, regardless of the details of how they are defined, their formation remains an open question. \cite{2018MPLA...3330004P} provide an extensive summary for the different ideas put forward as well as an assessment of the successes and failures of the proposed theories.
One such idea is that the satellites are beamed along filamentary structures of the cosmic web \citep[e.g. see][among others. Note that \citealt{2012MNRAS.424...80P} disputes this hypothesis on the basis that cosmic filaments are generally thicker than the scales under consideration; however such a hypothesis could work for the planes around Centaurus A which are considerably thicker than e.g. that around the MW]{1997MNRAS.290..411T,2004ApJ...603....7K,2011MNRAS.411.1525L, 2011MNRAS.413.3013L,2018MNRAS.476.1796S}. Indeed \cite{2015MNRAS.452.1052L} found evidence that the cosmic web may be at least partly responsible given that the normals of four dwarf galaxy planes are aligned with the axis of greatest compression of the velocity field as defined by the tidal tensor, \eone. However, some preferential alignment with the cosmic web might also be expected for the two most popular alternative formation scenarios. Both the  direction of infall of groups of dwarf galaxies, as well as the orbital orientation of major galaxy interactions that can result in the formation of tidal dwarf galaxies, can be expected to be oriented along the cosmic web. As such, it will be crucial to study the expected alignments for these proposed formation scenarios of dwarf galaxy planes in more detail.

In this paper we have extended the work of \cite{2015MNRAS.452.1052L}  in two important ways: we considered a more sophisticated measure of the cosmic web, known as the QL reconstruction \citep{2018NatAs...2..680H}, and we have examined the location of an additional 5 planes published in the literature. The alignments between plane normals and the tidal tensor's eigenvectors first published by \cite{2015MNRAS.452.1052L}  are recovered here. The additional 5 normals -- those belonging to the planes around  M83, M101, two non-satellite Local Group planes and the Great Northern plane, are not as well aligned with \eone~as the four planes around Centaurus A and M31. 

However, the fact that all the plane normals avoid the  \etwo~or \ethree~axes by $\gsim45$ degrees suggests that even the new, less correlated planes, are unlikely to be completely un-influenced by the eigenvectors of the tidal tensor. Any model of the formation of the planes should be able to account for this observation. Nevertheless, assuming that their planarity is a real feature, we suggest three reasons for the lack of a tight alignment with the tidal tensor's eigenvalues:

1. The planes in question may be incomplete. Given that some of the new planes considered here, e.g around M101 have significantly fewer dwarf galaxies then e.g. the plane of dwarfs surrounding the Milky Way, the estimation of their normal direction may be uncertain. Furthermore the distances considered here are typically determined by resolving stars at the tip of red giant branch. Such an estimate has an error of 5-10\%, which at large distances (e.g  M101 is located at $\approx7$Mpc) can translate into half a Mpc or more. In this scenario, as telescopic sensitivity improves, distance errors are reduced and deeper observations become available, our census of the regions in question will improve and change the direction of the estimated normal.

2. The QL tidal tensor eigenvectors may be incorrect. The QL density field, and its derived properties such as the tidal tensor, are based on constrained simulations and realizations of observations of the peculiar velocity of around 8,000 tracers from the CosmicFlows-2 survey \citep{2013AJ....146...86T}. As mentioned above, the constraints can have distance errors at the 5-10\% level, thereby tarnishing the QL field with inaccuracy. Accordingly as the CosmicFlows survey increases its scope and accuract, the QL field constructed from it may change. On the other hand there is (obviously) a finite number of galaxies in the Local Universe and it is unclear by how much the QL eigenvectors can change as the prospect of discovering new galaxies in the nearby Universe wanes. 

3. The planes in question may have different origins and thus correlate with different scales of the tidal field. As mentioned at the end of section 3.1, the nature of the QL reconstruction is that the eigenframe changes smoothly throughout the local volume. Yet the directionality of the planes changed more erratically. Accordingly the planes which do not align with the tidal tensor may have been formed via completely divergent processes. Among them tidal galaxies, group accretion or an as yet unknown mechanism may be responsible. Such an explanation would be favoured if it were eventually affirmed that these planes are completely different objects. Such an explanation would be disfavoured if future observations show that local galaxy planes are all similar objects with similar characteristics and formation mechanisms. In this case, uncorrelated alignments with the eigenvectors of the tidal tensor would disfavour the hypothesis that such planes are primarily formed by accretion along the principal axes of the cosmic web. Namely, it may be a coincidence that 4/10 of the planes considered here are so well aligned with \eone.

A fourth, much simpler solution for the misalignment of at least some proposed planes of dwarf galaxies in the literature is given by the possibility that they are false positives. Future deep surveys will test this by sampling more dwarf galaxies, hence giving better insights into the 3 dimensional distribution of the dwarf galaxies.

\section*{Acknowledgments}
NIL is indebted to RYP for inspiring enlightenment. NIL also acknowledges financial support of the Project IDEXLYON at the University of Lyon under the Investments for the Future Program (ANR-16-IDEX-0005). OM thanks the Swiss National Foundation for financial support and the hospitality of the people at AIP during his stay.  MSP acknowledges that support for this work was provided by NASA through Hubble Fellowship grant \#HST-HF2-51379.001-A. YH has been partially supported by the Israel Science Foundation (1358/18). HC acknowledge support from the Lyon Institute of Origins under grant ANR- 10-LABX-66. JS acknowledges support from the ``l'Or\'eal-UNESCO Pour les femmes et la Science'' and the ``Centre National d'\'etudes spatiales (CNES)'' postdoctoral fellowship programs. AK is supported by the {\it Ministerio de Econom\'ia y Competitividad} and the {\it Fondo Europeo de Desarrollo Regional} (MINECO/FEDER, UE) in Spain through grant AYA2015-63810-P and the Spanish Red Consolider MultiDark FPA2017-90566-REDC. He further thanks Cool Hawaii for the island mellow series.

\bibliography{Allrefs}
\end{document}